\documentclass[10pt,twocolumn,letterpaper]{article}

\usepackage{wacv}              

\usepackage{graphicx}
\usepackage{amsmath}
\usepackage{amssymb}
\usepackage{booktabs}

\usepackage{multirow}
\usepackage{multicol}
\usepackage{booktabs}
\usepackage{amsmath}
\usepackage{array}
\usepackage{lipsum}
\usepackage{tabularx} 
\usepackage{xcolor}

\usepackage{colortbl}%

\usepackage[pagebackref,breaklinks,colorlinks]{hyperref}

\usepackage[capitalize]{cleveref}
\crefname{section}{Sec.}{Secs.}
\Crefname{section}{Section}{Sections}
\Crefname{table}{Table}{Tables}
\crefname{table}{Tab.}{Tabs.}

\def\wacvPaperID{2050} 
\def\confName{WACV}
\def\confYear{2025}

\begin{document}

\title{Dual-Representation Interaction Driven Image Quality Assessment with Restoration Assistance}

\author{
Jingtong Yue$^{1}$\quad  
Xin Lin$^{1}$\quad  
Zijiu Yang$^1$\quad 
Chao Ren$^{1}$\footnotemark[2]\quad\\
$^1$Sichuan University \quad
}

\maketitle

\begin{abstract}
No-Reference Image Quality Assessment for distorted images has always been a challenging problem due to image content variance and distortion diversity. Previous IQA models mostly encode explicit single-quality features of synthetic images to obtain quality-aware representations for quality score prediction. However, performance decreases when facing real-world distortion and restored images from restoration models.
The reason is that they do not consider the degradation factors of the low-quality images adequately. 
%
%
To address this issue, we first introduce the DRI method to obtain degradation vectors and quality vectors of images, which separately model the degradation and quality information of low-quality images. 
After that, we add the restoration network to provide the 
MOS score predictor with degradation information. Then, we design the Representation-based Semantic Loss (RS Loss) to assist in enhancing effective interaction between representations. 
Extensive experimental results demonstrate that the proposed method performs favorably against existing state-of-the-art models on both synthetic and real-world datasets. The source code will be released at \url{https://github.com/Jingtong0527/DRI-IQA}.

\end{abstract}

\maketitle
\begin{figure}
    \centering
    \includegraphics[width=8cm]{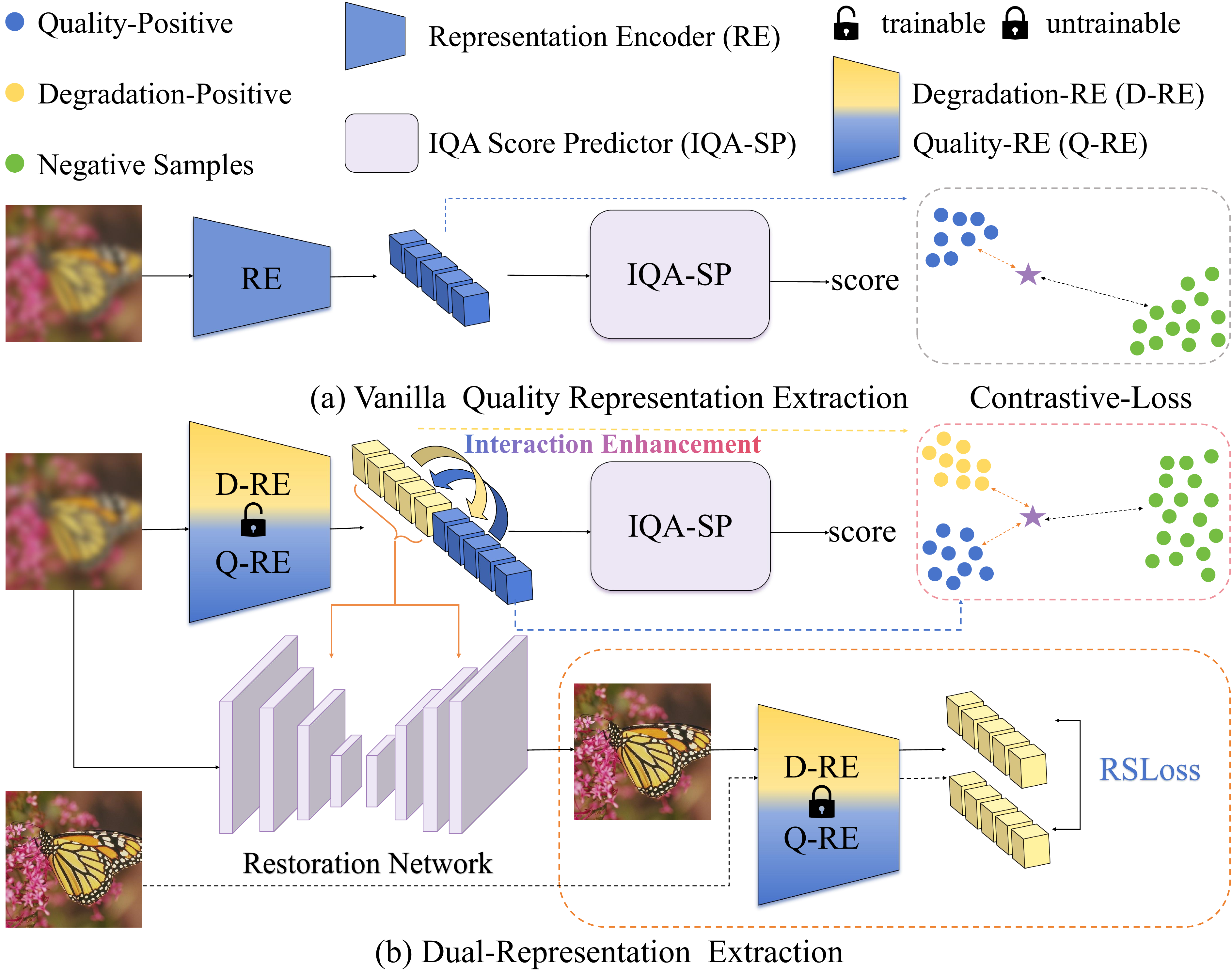}
    \caption{(a) The conventional pipeline \cite{QPT} to train a quality-aware representation encoder, the blue and green balls represent the Quality-Positive and Negative samples, respectively. (b) Our Dual-Representation Interaction method adds Degradation-Positive samples symbolized by yellow balls. We also propose Representation-based Semantic Loss (RS Loss) to constraint model training.
}
    \label{fig:intro}
\vspace{-15pt}
\end{figure}

\section{Introduction}
\label{1}

High-quality images are subject to various disturbances during compression, transmission \cite{metaiqa}, or shooting. These distortions lead to blurring, noise, color deviation, compression loss, and other problems that seriously affect the visual experience. Therefore, a method for evaluating the degree of image quality is particularly important. Full-Reference Image Quality Assessment (FR-IQA) methods is proposed to predict image quality score, which leverages reference images during testing. However, in real-world scenarios, FR-IQA models cannot work without reference images.
Thus, No-Reference Image Quality Assessment (NR-IQA) \cite{nriqa1, nriqa2,nriqa3} has been a popular topic recently as a promising approach to evaluate low-quality images without reference images. 

Meanwhile, numerous Image Restoration (IR) algorithms are proposed to restore low-quality images to its clean version effectively,
%
including GAN-based models \cite{gan1, gan2, scpgabnet, scpgan}, classical encoder-decoder networks \cite{nafnet}, and query-based transformer methods. 
Therefore, with the development and application of restoration networks, robust IQA models are required not only to evaluate the quality of original low-quality images but also to assess the quality of restored images accurately.
%
%
Some image restoration networks focus on local feature modeling \cite{local1, local2, local3}, leveraging the redundant information in receptive patterns to restore the images, while efforts \cite{bert, globe} are also made to model non-local and global image priors. However, the above-mentioned feature extraction strategies cause a decline in the generalization ability of IQA models. They may result in detail or overall quality losses in the restored image, which poses a significant challenge for IQA methods to accurately evaluate the visual quality of the restored images from different IR networks.

%
To tackle the above challenges, QPT extracts quality-aware representation to predict quality scores as shown in Figure~\ref{fig:intro}(a).
It first leverages contrastive learning with more mixed degradations added to the clean image to construct negative samples. After that, these samples are used to enhance the encoder's ability to perceive the overall quality feature of the image and weaken the sensitivity to a single degradation type, thus constructing a quality-aware representation of the quality score perception. 
However, it has limitations as follows:
(1) Integrating image content and degradation factors together leads to insufficient attention to the degradation factors, which has a significant impact on image quality.
(2) The impact of restoration model characteristics on the image quality assessment cannot be ignored.
As stated in \cite{desra}, the restored results of the GAN-based approaches \cite{gan1, gan2} have obvious artifacts, and these regions are considered high-quality by a quality-aware predictor. In order to improve the performance of IQA models for restoration results \cite{CKDN, reiqa}, guidance from the restoration model is necessary.

Based on the above discussion, we propose the Dual-Representation Interaction (DRI) method, which utilizes representation interaction between quality-aware and degradation-aware, and restoration network assistance to improve the robustness of NR-IQA models. The detail is shown in Figure~\ref{fig:intro}(b). Specifically, we offer image degradation-related representations to the IQA model through the implicit interactions of different regions in one representation. After that, to enhance the generalization of the proposed model across different restoration networks and distortion types, we utilize restoration networks as an assistance. In addition, to further enhance the interaction between the two types of representations, we design a Representation-based Semantic loss (RS Loss), which is a penalty mechanism between the degraded representations of the restored image and the clean image.

Extensive experimental results demonstrate that our DRI-IQA performs comparably to the existing SOTA methods on various datasets, including synthetically distorted, real-world degraded, and GAN-based restored datasets.  Our contributions are as follows:

\begin{itemize}
\item  
We propose the Dual-Representation Interaction (DRI) method to obtain representations with different attributes that promote each other through latent interactions for high-performance NR-IQA.

\item  We design a Restoration-Assistance Module (RAM), which provides the IQA model with quality-related degradation details and a trainable Representation-based Semantic Loss (RS-Loss) to enhance the quality-related degradation extraction and effective interactions between the representations.
\item  We explore the strong correlation between restoration tasks and IQA tasks, and implicitly connect the two tasks within the restoration network through representation embedding, creating a promising restoration-assisted IQA framework.
\end{itemize}
\begin{figure*}[ht]
\includegraphics[width=1\linewidth]{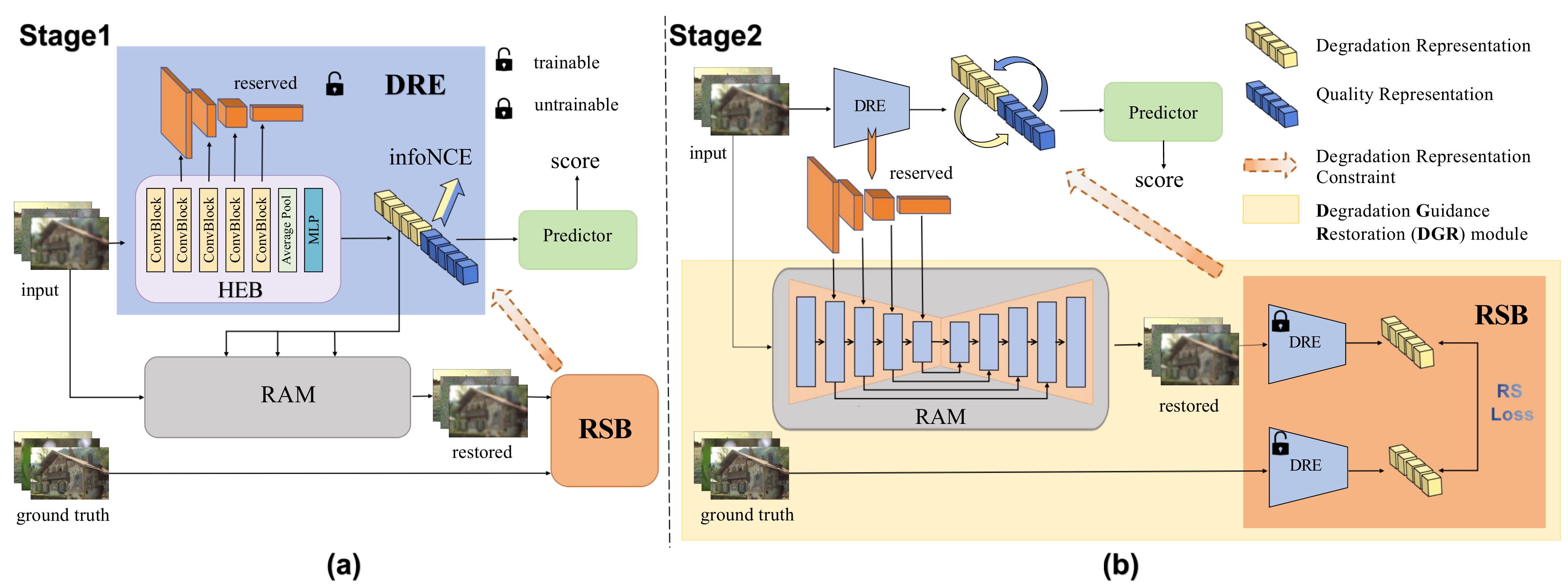}
\caption{\label{nerwork}
Overview of our two-stage Dual-Representation Interaction IQA (DRI-IQA) architecture. Stage 1 is set to train the Dual-Representation Extractor Net (DRE-Net), extracting quality representation and degradation representation simultaneously and reserving the features for guiding the Restoration Network. The modules in the blue region are the components we use in stage 1. In Stage 2, we train the score predictor with the assistance of the Restoration Network, that is the complete network joins in training. Meanwhile, the Representation-based Semantic Loss (RS Loss) is leveraged to promote implicit interaction between two representations and enhance the performance of the Restoration Network.}
\vspace{-15pt}
\end{figure*}
\section{Related Work}

\subsection{No-Reference Image Quality Assessment}

No-Reference Image Quality Assessment (NR-IQA) aims to predict the quality score of low-quality images without reference images. However, the absence of a reference image also makes image quality evaluation difficult, especially for real-world distorted images. Previous works \cite{NSS2, NSS3, NSS4, NSS5} extract features in transformed domains with Natural Scene Statistics (NSS) method \cite{NSSo}, which utilize parameters derived from fitting a Gaussian model as features of distorted images. For instance, the famous and widely used metric Natural Image Quality Evaluator (NIQE) \cite{niqe} is NSS-based and leverages features that capture in a normalized band-pass space. Recently, with the success of n Convolutional Neural Networks (CNN) \cite{vgg, alexnet, resnet} and Transformer \cite{vit1, vit2, vit3, vit4, vit5, vit6}, the performance of the NR-IQA algorithm is significantly improved. 

For CNN-based models, CKDN \cite{CKDN} exploits the degradation of restored images from IR models and extracts reference information from degraded images by distilling knowledge from pristine-quality images. Hyper-IQA \cite{hyperiqa} establishes perception rule by a hyper network and predicts image quality in a self-adaptive manner. 

For the Transformer backbone, MUSIQ \cite{MUSIQ} captures image quality at different granularities with a multi-scale image representation, which accommodates the problem of fixed shape
constraint in CNN. MANIQA \cite{MANIQA} firstly extracts features via Vision Transformer (ViT), then strengthens global and local interactions aiming to improve the performance on GAN-based distortion images. However, the methods above mostly focus on single-quality features and overlook the degradation corresponds. Our approach will address this issue and generalize on common distorted and IR network restored images.

\subsection{Multi-degradation Image Restoration}

The image Quality Assessment (IQA) model is usually used to evaluate synthetic and real-world distorted images. However, with the development of Image Restoration (IR) methods, IQA models are also leveraged to evaluate their effectiveness. Current IR methods \cite{scpgabnet, deblur, lin2024dual, liao2025asymmetric, pan2022real} focus on one single type of degraded images, but for more practical reasons, people prefer to design an all-in-one model that supports restoring multiple degraded or mixed degraded images in one model. For instance, the Ingredients-oriented Degradation Reformulation framework \cite{ingredient} conducts ad hoc operations on different degradations according to the underlying physics principles and employs a dynamic routing mechanism for probabilistic unknown degradation removal. Moreover, AMIRNet \cite{amirnet} learns a degradation representation for unknown degraded images by progressively constructing a tree structure through clustering, without any prior knowledge of degradation information.

Besides, many works also incorporate contrastive learning \cite{contrastive} to extract multiple degradation features and boost model generalization. Examples of this can be found in DASR \cite{dasr} and AirNet \cite{AirNet}. DASR proposes an unsupervised degradation representation learning scheme for blind super-resolution based on a contrastive learning strategy. AirNet is the first to propose an all-in-one IR network, which could recover various degraded images in one network.

The aforementioned works show that the degraded features learned by the classic IR encoders may be valuable and meaningful to IQA tasks. Therefore, we propose the IR network assistance strategy which enables the IQA model to address the issue of poor generalization to various distorted images. 
\subsection{Loss Function For Image Restoration}
The loss function serves as a tool to constrain the model's training process and is critical to the model's performance. Nowadays, most image restoration models use L1 and L2 loss as constraints for training. They are pixel-level constraints designed to bring the restored and clean ground truth images closer together on a pixel-by-pixel basis from a mathematical level. Moreover, L1 and L2 focus on the similarity of the two images and are easily disturbed by the content details. Perceptual Loss \cite{perceptual} focuses on the human eye's perception of an image, utilizing pre-trained VGG on ImageNet to extract global features from recovered and corresponding clean images and constrain them at the feature level. However, the datasets used for training the VGG model are high-quality images, without prior features about image quality variations and degradation information. To address this issue, we introduce the Representation-based Semantic Loss (RS Loss), which focuses on the similarity of their features related to the degradation of quality, and tends to make the degradation of the recovered image similar to that of the clean image in the penalization process. Meanwhile, it assists in learning degradation representation, incorporating features from the restoration network, and enhancing the performance of the IQA model.
\section{Proposed Methods}
In this section, we first introduce the proposed Dual-Representation Interaction IQA (DRI-IQA) (\ref{3.1}), which mainly includes the Dual-Representation Extractor (DRE) (\ref{3.2}) and the Restoration-Assistance Module (RAM)(\ref{3.3}). Next, the Representation-based Semantic Loss (RS-Loss) is presented (\ref{3.4}).


\subsection{Dual-Representation Interaction (DRI)}
\label{3.1}

We propose a Dual-Representation Interaction (DRI) method to obtain a dual-representation (quality-degradation) and enhance the latent interaction to compensate for the quality-aware degradation details for IQA tasks. 
As shown in Figure~\ref{nerwork}(a), we pre-train the DRE by infoNCE Loss \cite{moco} in the stage 1. 
Specifically, we take a batch of low-quality images with hybrid degradation as input to the Hybrid Encoder with five layers to acquire a representation. Then, we split the obtained representation and constrain the upper half to be degradation-aware and the lower half to be quality-aware by the corresponding loss function.
To adapt to various degradation types, the DRI method adopts a randomized hybrid degradation method to construct image pairs dynamically for pre-training DRE. 
Specifically, we use a combination of the DIV2K \cite{div2k} and Flickr2k \cite{flicker} to construct a clean image dataset and synthesize the corresponding low-quality images by the following approach.
Assume that the degradation function is \(D_{\text{i}}(x, \omega_{\text{i}}, p_{\text{i}})\), where \(\omega\) refers to the degree coefficient of each degradation and \(p\) denotes the probability of randomly selecting each degradation.

In the training process, we take a clean image \(x\in R^{\text{C$\times$H$\times$W}}\) and randomly synthesize multiple degradation twice to obtain \(x_{\text{1}}\in R^{\text{C$\times$H$\times$W}}\) and \(x_{\text{2}}\in R^{\text{C$\times$H$\times$W}}\). The images undergo multiple iterations to obtain degradations with random orders, types, and degrees. The initial iteration is defined as \(x_{\text{1,1}} =  D_{\text{1}}(x,\omega_{\text{1,1}},p_{\text{1,1}})\) and \(x_{\text{2,1}} =  D_{\text{1}}(x,\omega_{\text{2,1}},p_{\text{2,1}})\), then iterations are performed according to the following formula:
\begin{equation}\label{eq:x1_d}
\left\{
\begin{aligned}
\begin{split}
    &x_{\text{1,i}} =D_{\text{i}}(x_{\text{1,i-1}},\omega_{\text{i,1}},p_{\text{i,1}})\\
    &x_{\text{2,i}} =  D_{\text{i}}(x_{\text{2,i-1}},\omega_{\text{i,2}},p_{\text{i,2}})
    \end{split}
    \quad
    , 1 \leq i \leq N,
\end{aligned}
\right.
\end{equation}
where $N$ represents the number of times adding degradation. Due to the presence of skip operation, the value of n varies between different batches, but it will not exceed 6 times at maximum.

During stage 2 in Figure~\ref{nerwork}(b), we put the low-quality images to the DRE first, acquiring reserved features and dual-representation. After that, the quality representation is used to predict quality scores, and the degradation representation goes through RAM. Finally, in the Representation-based Semantic Block (RSB), we put the restored image and high-quality ground truth together into the fixed DRE and constrain the output degradation representation by the proposed Representation-based Semantic Loss (RS Loss).

\begin{figure}
    \centering
    \includegraphics[width=8.5cm]{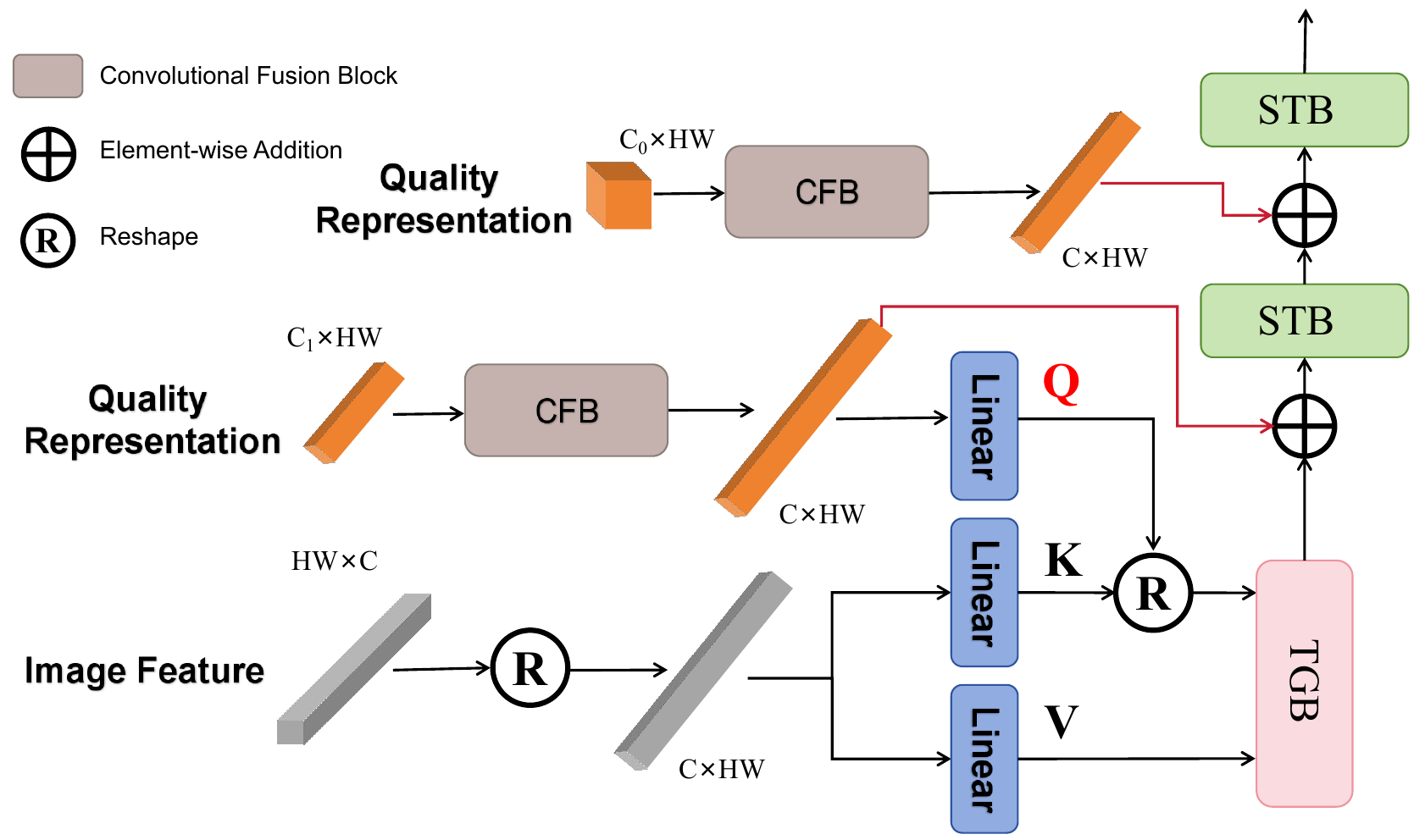}
    \caption{Details of our guidance strategy. The cuboid in orange and gray represent the quality-aware part of Dual-Representation and the features encoded from input images, respectively. Also, the direction of the cuboid represents the transposed form. TGB denotes Transposed Guidance Block and STB represents Swin Transformer Block \cite{MANIQA}. 
}
\label{fig:guide}
\end{figure}

For the MOS score prediction, the images and representations are together sent to the Transformer-based IQA network \cite{MANIQA}. As shown in Figure \ref{fig:guide}, we set the image features as K, V, and the quality representations as Q in the Transformer block to guide the output of the final score. In addition, we add the Convolutional Fusion Block (CFB) consisting of convolutional layers and concatenation operation, allowing the number of channels in the quality representation \(C_{\text{0}}\) and \(C_{\text{1}}\) to match the number of channels in the image features \(C\). 
\begin{figure}
    \centering
    \includegraphics[width=8.3cm]{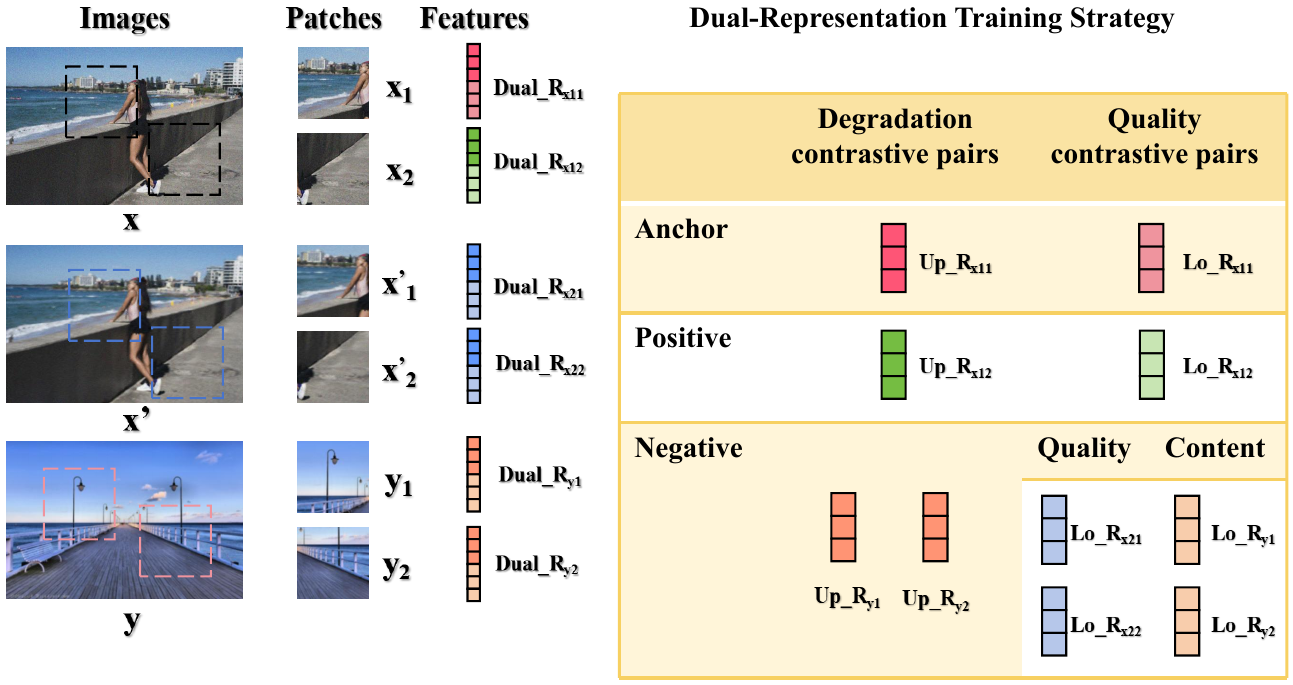}
    \caption{The \( x_{\text{1}}\) and \(x_{\text{2}}\) are generated from the same input images with varied degradation distortion, while \( y \) is the input image with both different content and distortion. We get two patches randomly on each of the three images and put them into the Dual-Representation Extractor (DRE) to obtain the features. The table shows the choice strategy for the positive and negative samples of the two halves of the features.}
    \label{fig:rep}
\end{figure}


\subsection{Dual-Representation Extractor (DRE)}
\label{3.2}
To allow degraded information to supplement quality information, we leverage one encoder to extract the dual-representation and promote the interaction between the degradation-aware region and the quality-aware region.

Specifically, as illustrated in Figure \ref{fig:rep}, a random crop on the image  \(x_{\text{1}}\) named \(x_{\text{11}}\in R^{\text{C$\times$H$\times$W}}\)and \(x_{\text{12}}\in R^{\text{C$\times$H$\times$W}}\), respectively. Take the \(x_{\text{11}}\) shown in ~\ref{fig:rep} for an example, it is then fed into the Hybrid Encoder \(Enc(\bullet)\), obtaining a long representation \(Dual\_R_{\text{x11}} \in R^{\text{1024$\times$H$\times$W}}\). Then, take the upper half, we get \(Up\_R_{\text{x11}}\in Dual\_R_{\text{x11}}\), as a degradation representation, and we make it as the Anchor, another \(Up\_R_{\text{x12}}\in R^{\text{512$\times$H$\times$W}}\)as the positive sample, then let all the remaining patches that are not only semantically different from \(x_{\text{1}}\) but also have different degradation types as the negative samples (\(Up\_R_{\text{y1}}\),\(Up\_R_{\text{y2}}\)), and the final degradation information modeling results are obtained by using infoNCE Loss \cite{moco} for constraints.

On the other hand, the lower part of \(Dual\_R_{\text{x11}}\), assumed as  \(Lo\_R_{\text{x11}} \in Dual\_R\), is used as an Anchor of quality representation. Compared to the negative samples of degradation part, the quality portion adds the \(Lo\_R_{\text{x21}}\in R^{\text{512$\times$H$\times$W}}\) and \(Lo\_R_{\text{x22}}\in R^{\text{512$\times$H$\times$W}}\), which are obtained by cropping and encoding via \(x_{\text{2}}\). Thus, images with the same content but different degrees of quality can also be learned, reducing the interference of background information on quality modeling.
The experimental results show that even without using large-scale datasets such as ImageNet for pre-training, the small-scale high-quality dataset combined with dual-representation can achieve great IQA performance, which greatly saves computational resources and training time.

\subsection{Restoration Assistance Module (RAM)}
\label{3.3}
As we discussed in Section \ref{1}, the appropriate guidance of degradation information is important to improve the accuracy and generalization of the IQA algorithm. Therefore, we feed the distorted image into the restoration network NAFNet~\cite{nafnet} in the RAM module and use the reserved features to guide the encoder and decoder of the NAFNet layer by layer to realize the restoration of images.

This has two benefits: first, through the constraints of the restoration loss, the DRE module can better extract degradation features of the low-quality image x. Secondly, the bootstrapping process can enable DRE to learn certain restored features of the restoration network, which enhances the performance of the same IQA algorithm in evaluating different types of restored images. Since both representations stem from the same encoder simultaneously, the above two points can influence the quality representation through implicit interaction, thereby enhancing the performance of the IQA model.
\begin{equation}\label{eq:restore}
x_{r} = RAM(x, Up\_R),
\end{equation}
\begin{equation}\label{eq:resloss}
Loss = ||Ref-x_{r}||_{\text{2}} + \lambda Perceptual(Ref, x_{r}),
\end{equation}
where \textit{Ref} represents pristine-quality images and \textit{Perceptual} refers to the perceptual loss \cite{perceptual}. Moreover, \( \lambda\) is set to be 0.01.
\subsection{Representation-based Semantic Loss}
\label{3.4}

In our DRI-IQA framework, the role of the restoration network is to assist the IQA model in accurately evaluating the quality of different types of restored images. However, only utilizing degradation to guide the NAFNet, constraining with L2 Loss and Perceptual Loss as shown in equation~\ref{eq:resloss} is insufficient. The L2 Loss is a pixel-level constraint, and the Perceptual Loss uses the pre-trained VGG model, which focuses only on semantic information without degradation-related representation.


Therefore, to adequately learn the characteristics of the restoration network and effectively interact with the quality representation, we design a Representation-based learning Semantic loss function (RS Loss). We put the restored image obtained by NAFNet together with the ground truth images into the DRE module to obtain the respective degradation representation and then constrain by the Representation-based Semantic loss:
\begin{equation}\label{eq:halfres}
Up\_R_{\text{res}} = Half(DRE(Restored)),
\end{equation}

\begin{equation}\label{eq:halfref}
Up\_R_{\text{ref}} = Half(DRE(Ref)), 
\end{equation}

\begin{equation}\label{eq:rsloss}
L_{\text{RS}} = ||Up\_R_{\text{ref}} - Up\_R_{\text{res}}||_{\text{2}},
\end{equation}
where \(Up\_R_{\text{res}}\) and \(Up\_R_{\text{ref}}\) denote the degradation representation of the restored image and the high-quality image, and the $Half(\bullet)$ function serves to separate the original representation by taking the upper half, \(||\bullet ||_{\text{2}}\) denotes the conventional loss function L2 Loss for the image level.

\begin{figure*}[ht]\label{fig:mos}
\includegraphics[width=1\linewidth]{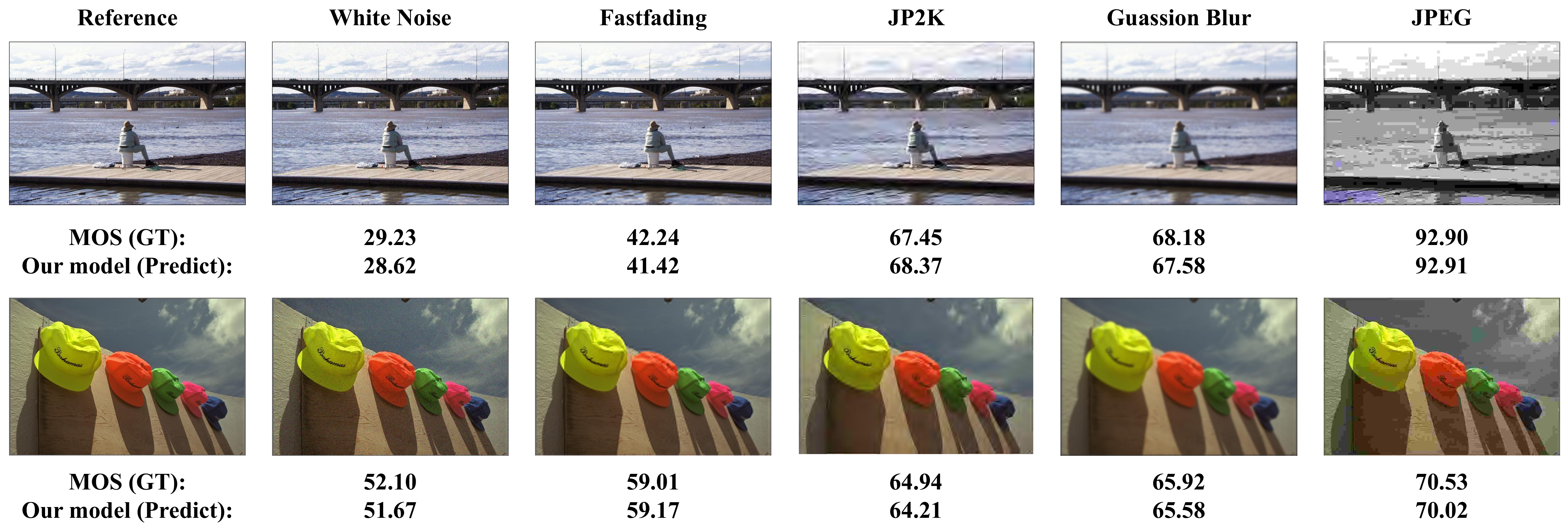}
\caption{\label{mos}Predicted MOS scores by our DRI-IQA model and the MOS score labels on low-quality images with varying degradations.}
\end{figure*}

\begin{table}
\fontsize{8}{8}\selectfont
\caption{The summary of used Datasets.} 
\setlength{\tabcolsep}{1mm}{
\begin{tabular}{cccc}
\toprule
Dataset & \# of Dist.Images &  \# of Dist.Types & \# of Ref.Images \\
\midrule
LIVE\cite{live} & 779 & 5 & 29\\
\midrule
KADID-10K\cite{kadid} & 10,125 & 125 & 81   \\
\midrule
KonIQ-10k\cite{koniq} & 10,073 & authentically & -   \\
\midrule
LIVEC\cite{clive}   & 1162 & authentically & -  \\
\midrule
PIPAL\cite{pipal}      & 29,000 & 40 & 200    \\
\midrule
\end{tabular}}%
\label{dataset} 
\end{table}%

\begin{table*}
\centering
\fontsize{10}{10}\selectfont
\caption{Image Quality Assessment results of DRI-IQA on four standard datasets. Some results are from previous works \cite{MANIQA, QPT, reiqa}.} 
\setlength{\tabcolsep}{2mm}{
\begin{tabular}{ccccccccccc}
\cmidrule{1-10}
\multirow{2}[3]{*}{Methods} & \multirow{2}[3]{*}{Conference/Journal}&  \multicolumn{2}{c}{LIVE} &  \multicolumn{2}{c}{KADID-10K} &  \multicolumn{2}{c}{KonIQ-10k} &  \multicolumn{2}{c}{CLIVE}  \\
\cmidrule{3-10}
&&  SROCC & PLCC & SROCC & PLCC & SROCC & PLCC& SROCC & PLCC \\
\cmidrule{1-10}
\cmidrule{1-10}
BRISQUE \cite{brisque}  &TIP2002& 0.939 & 0.935 & 0.528 & 0.567 & 0.665 & 0.681 & 0.608 & 0.629 & \\
\cmidrule{1-10} 

CORNIA \cite{cornia}  &CVPR2012 & 0.947 & 0.950 & 0.516 & 0.558 & 0.780 & 0.795 & 0.629 & 0.671 &  \\
\cmidrule{1-10}
\cmidrule{1-10}
DB-CNN \cite{dbcnn}   &TCSVT2018& 0.968 & 0.971 & 0.851 & 0.856 & 0.875 & 0.884 & 0.851 & 0.869 &  \\
\cmidrule{1-10}

PQR \cite{pqr}     &CVPR2017 & 0.965 & 0.971 & -     & -     & 0.880 & 0.884 & 0.857 & 0.882 &  \\
\cmidrule{1-10}

PaQ-2-PiQ \cite{p2p}&CVPR2019& -     & -     & -     & -     & 0.870 & 0.880 & 0.840 & 0.850 &  \\
\cmidrule{1-10}

HyperIQA \cite{hyperiqa} &CVPR2020& 0.962 & 0.966 & 0.852 & 0.845 & 0.906 & 0.917 & 0.859 & 0.882 &  \\
\cmidrule{1-10}
CONTRIQUE \cite{CONTRIQUE}&TIP2021& 0.960 & 0.961 & 0.934 & 0.937 & 0.894 & 0.906 & 0.845 & 0.857 &  \\
\cmidrule{1-10}
QPT \cite{QPT}   &CVPR2023& - & - & - & - & 0.927 & 0.941 & 0.895 & 0.914  \\

\cmidrule{1-10}
\cmidrule{1-10}
\cmidrule{1-10}
MUSIQ \cite{MUSIQ}&ICCV2021& -     & -     & -     & -     & 0.916 & 0.928 & -     & -     &  \\
\cmidrule{1-10}

TRes \cite{tres}   &WACV2022& 0.969 & 0.968 & 0.859 & 0.858 & 0.915 & 0.928 & 0.846 & 0.877  \\
\cmidrule{1-10}
\cmidrule{1-10}
Re-IQA \cite{reiqa}   &CVPR2023& 0.970 & 0.971 & 0.872 & 0.885 & 0.914 & 0.923 & 0.840 & 0.854 &  \\
\cmidrule{1-10}
DRI-IQA (Ours) &WACV2025& 0.982 & 0.984 & 0.941 & 0.943 & 0.936 & 0.948 & 0.861 & 0.886 \\
\cmidrule{1-10}

\end{tabular}}%
\label{all}
\end{table*}
\begin{figure}
    \centering
    \includegraphics[width=8cm]{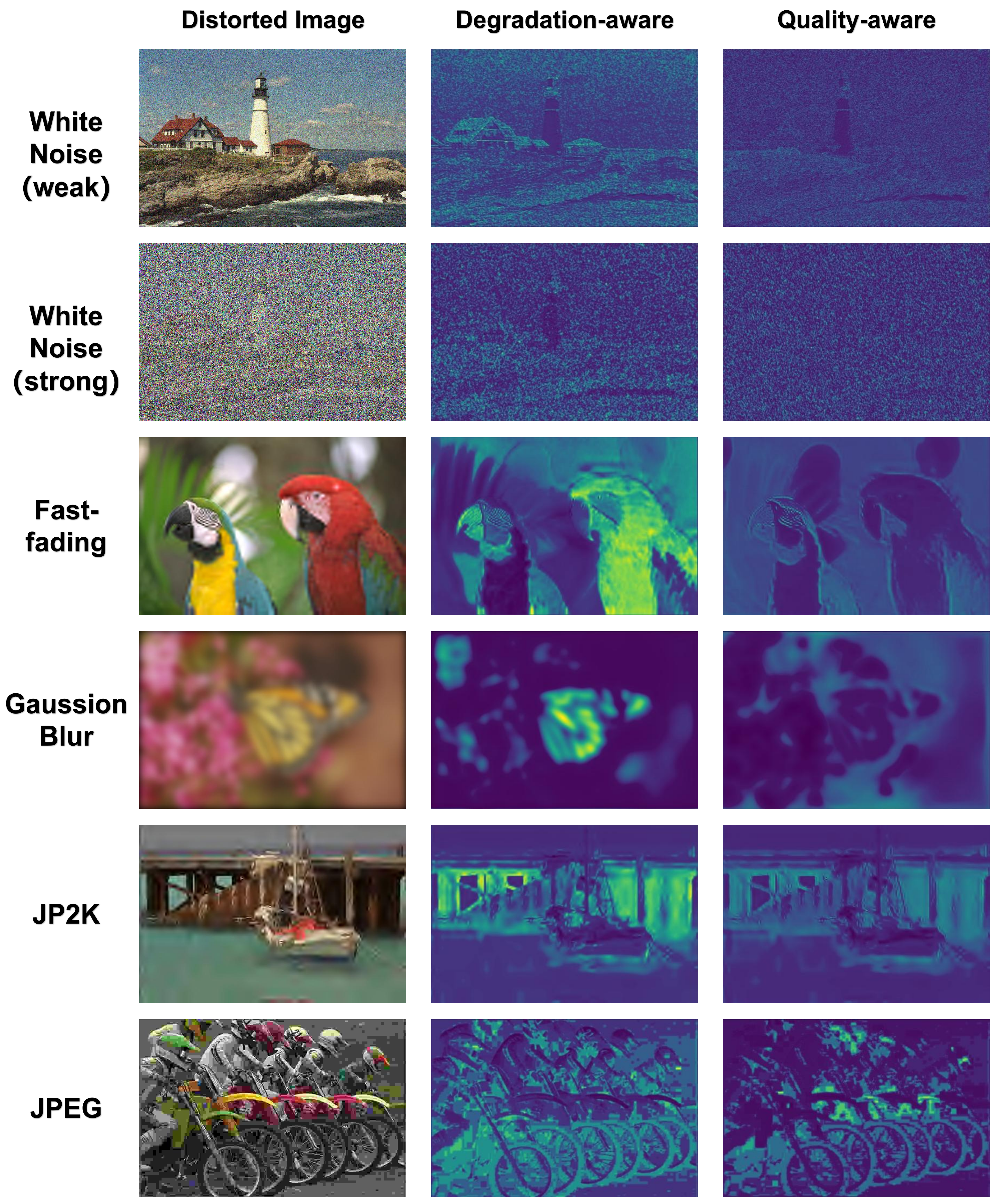}
    \caption{The feature maps of quality representation and degradation representation extracted by DRE from low-quality images.
}
\label{fea_map}
\vspace{-15pt}
\end{figure}
\section{Experiment}


In this section, we first describe the dataset we used for training/testing and present the implementation details. Then, we provide some qualitative and quantitative comparison results with the existing methods. Finally, extensive ablation studies validate the effectiveness of the proposed DRI-IQA framework.

\subsection{Datasets}
\textbf{Pre-Training Datasets.} We randomly select 3392 images from the widely used DIV2K \cite{div2k} and Flicker2K \cite{flicker} for pre-training our Dual-Representation Extractor (DRE).\\
\textbf{IQA Datasets.} We train and test our DRI-IQA on two synthetic and two real-world IQA datasets. For the experiments on synthetic data, we use the LIVE \cite{live} and KADID \cite{kadid} datasets, where low-quality images are synthesized from their corresponding version. The LIVE dataset comprises 29 high-resolution and high-quality color images as reference images. These reference images undergo degradation using five types of computer distortion operations, including Gaussian blur, JPEG compression, JPEG2000 compression, JPEG2000 fast scale fading distortion, and white noise, resulting in 779 distorted images across 5 to 6 degradation levels each. The KADID-10K dataset comprises 81 pristine-quality images, each subjected to 25 instances of distortion degradation across 5 levels, resulting in a total of 10,125 low-quality images.

For the real-world scenes, we employ KonIQ-10k \cite{koniq} and LIVEC \cite{clive} datasets. KonIQ-10k \cite{koniq} is a large-scale IQA dataset comprising 10,073 quality-scored images. This is the first in-the-wild database aiming for ecological validity, with regard to the authenticity of distortions, the diversity of content, and quality-related indicators. Known as the ``Wild Image Quality Challenge Dataset'',  LIVEC \cite{clive} contains 1,162 images captured using modern mobile devices in a wide variety of real-world distortion scenarios from different photographers using different camera equipment in the real world, and thus these images contain complex real-world distortions. The subjective scores are collected through crowd-sourcing and evaluated by more than 8,100 testers, collecting more than 350,000 subjective opinion scores on the 1,162 images.

We also conduct experiments on the PIPAL \cite{pipal} dataset, which contains images processed by image restoration and enhancement models (particularly the deep learning-based methods) besides the traditional distorting methods. The dataset contains 29k images in total, including 250 high-quality reference images. It is challenging for existing metrics to predict perceptual quality accurately, especially the images generated by GAN-based IR methods. More details are shown in Table~\ref{dataset}. 
\vspace{-0.5em}
\subsection{Implementation Details}
In stage 1, we pre-train the Dual-Representation Extractor (DRE) following the configuration proposed in MoCo \cite{moco} for 300 epochs. Next, we train the MOS score predictor for 200 epochs on all datasets. All experiments are conducted on two Nvidia GeForce 2080Ti GPUs with Pytorch 1.11.0 and CUDA 11.3 for training and testing.


In the stage of pre-training, we set the learning rate l to \(2\times10^{\text{-4}}\) initially. The batch size B is set to 64, and we utilize the ADAM optimizer with weight decay \(1\times10^{\text{-8}}\) and cosine annealing learning rate with the parameters Tmax and etamin set to 300 and \(1\times10^{\text{-6}}\). 
For the quality score predictor, the batch size B is set to 16 and we utilized the ADAM optimizer with weight decay \(1\times10^{\text{-8}}\) and cosine annealing learning rate with the parameters Tmax and etamin set to 300 and \(1\times10^{\text{-6}}\). 
%
For the training loss, we leverage Mean Square Error (MSE) Loss. In addition, InfoNCE Loss \cite{moco} is used to constrain DRE-Net in both Stage 1 and Stage 2 during training. 

During testing, we randomly crop 224 $\times$ 224 patches 25 times from the original one. We follow the previous work \cite{MANIQA} and \cite{QPT} to generate the quality score by predicting the mean score of these 25 patches, and all results are averaged by 10 times split. The experiment is conducted 5 times with different seeds, and all 5 results are averaged to get the final results.
\subsection{Evaluation Metrics}
For the evaluation metrics, we follow the previous works using Spearman’s rank order correlation coefficient (SROCC) and Pearson’s linear correlation coefficient (PLCC) to evaluate the performance of our models. SROCC is used to measure the monotonicity of the IQA algorithm predicted score, the equation of which is defined as:

\begin{equation}\label{eq:srocc}
SROCC = 1 - \frac{6\sum_{i=1}^{N}d_{\text{i}}^{\text{2}}}{N(N^{\text{2}}-1)},
\end{equation}
where \(d_{\text{i}}\) denotes the difference between the subjective quality score ranking and the objective quality score ranking of the \(i^{\text{th}}\) image, and $N$ denotes the number of samples. The \(s_{\text{i}}\) and \(p_{\text{i}}\) denote the subjective and objective quality scores of the \(i^{\text{th}}\) image, respectively. The \(\bar{s}\) and \(\bar{p}\) denote the average subjective and objective quality scores, respectively, the equation of PLCC as follows:
\begin{equation}\label{eq:plcc}
PLCC = \frac{\sum_{i=1}^{N}(s_{\text{i}}-\bar{s})(p_{\text{i}}-\bar{p})}{\sqrt{\sum_{i=1}^{N}(s_{\text{i}}-\bar{s})^{\text{2}}\sum_{i=1}^{N}(p_{\text{i}}-\bar{p})^{\text{2}}}}.
\end{equation}
PLCC is used to assess the accuracy of IQA model predictions. Both SROCC and PLCC are in the range of [-1,1]; the closer they are to 1, the better the IQA model performs.

\subsection{Evaluated Methods}
We select two classical approaches, BRISQUE \cite{brisque} and CORNIA \cite{cornia}, which rely on hand-crafted features to assess image quality, and have been widely used in traditional no-reference image quality assessment (NR-IQA) tasks. 
Next, we incorporate a set of CNN-based quality score predictors, which represent more advanced techniques compared to hand-crafted approaches. These methods leverage deep convolutional networks to automatically learn relevant features from data, offering more robust and accurate predictions. For our experiments, we include DB-CNN \cite{dbcnn}, and Re-IQA \cite{reiqa}. 
In addition to the CNN-based models, we compare our proposed method, DRI-IQA, with the current state-of-the-art Transformer-based models, which have recently gained significant attention in the field of image quality assessment. These models use attention mechanisms to capture global dependencies and model long-range interactions between image patches more effectively. Specifically, we evaluate MUSIQ \cite{MUSIQ} and TReS \cite{tres}. 

\subsection{Image Quality Assessment Results}
Table~\ref{all} shows the quantitative evaluation of the proposed DRI-IQA model and other state-of-the-art image quality assessment methods. For CNN-based IQA methods, DRI-IQA gains 0.004, 0.017, 0.025 of SROCC in the KoiIQ-10k dataset compared to QPT \cite{QPT}, Re-IQA \cite{reiqa} and CONTRIQUE \cite{CONTRIQUE}.

For Transfomer-based IQA methods, our DRI-IQA achieves the best results across multiple test sets. Specifically, DRI-IQA yields improved performance by 0.082 of SROCC and 0.085 of PLCC in the KADID-10K dataset compared to the TRes \cite{tres}. Furthermore, in the KonIQ-10k dataset, DRI-IQA gains 0.021, 0.02 of SROCC and 0.02, 0.02 of PLCC compared to  TRes \cite{tres} and MUSIQ \cite{MUSIQ}. For the perceptual results, the comparison between the ground truth MOS score and the MOS score predicted by the proposed model is shown in Figure~\ref{mos}. The distorted images are arranged in descending order of quality. It is noticeable that different degradation types in the images do not affect the distinct visual differences between them. Moreover, the DRI-IQA model's predicted scores highly correlate with both visual perceptions and ground truth scores.
\subsection{Ablation Study}
In this section, we present the results of our ablation experiments on multiple datasets.
{\flushleft \textbf{Effectiveness of the proposed modules.}} Table~\ref{viqa} shows the ablation results on PIPAL and KonIQA-10k datasets. The V1, V2, and V3 represent a single quality-aware representation guidance network, dual-representation guidance and dual-representation guidance with restoration assistance, respectively. It is clear that the quality score predictor with only DRI method (V2) gains 0.03 of SROCC and 0.06 of PLCC in the PIPAL dataset compared to the score predictor guided by single quality representation (V1). This verifies that the Dual-Representation Interaction (DRI) method is essential in improving the performance of the IQA model. 

Meanwhile, V3 gains 0.003 of SROCC and 0.003 of PLCC compared to V2, demonstrating the effectiveness of restoration assistance. Furthermore, the proposed method improves performance by 0.002 of SROCC and 0.007 Of PLCC in the PIPAL dataset, which verifies the effect of RS Loss. 

In addition, Figure~\ref{fea_map} clearly shows that degradation information is more sensitive to different degradation types, while quality representation is less affected by semantic interference, due to the inclusion of negative samples with the same content but different degradation types, making it more conducive to quality assessment.\\
\begin{table}
\fontsize{10}{10}\selectfont
\caption{Ablation study on proposed modules. V1 = single quality-aware representation + score predictor; V2 = Dual-Representation + score predictor; V3 = V2 + NAFNet; Proposed = V3 + RS Loss.} 
\setlength{\tabcolsep}{3mm}{
\begin{tabular}{ccccc}
\toprule
& \multicolumn{2}{c}{PIPAL} &\multicolumn{2}{c}{KonIQ-10k}  \\
\cmidrule{2-5}
 & SROCC & PLCC & SROCC & PLCC  \\
\midrule
V1  & 0.690 & 0.714 & 0.931 & 0.945   \\
\midrule
V2 & 0.693 & 0.720 & 0.934  & 0.947 \\
\midrule
V3 & 0.696 & 0.723 & 0.934  &0.948 \\
\midrule
Proposed & 0.698 & 0.730 & 0.936 & 0.948  \\

\midrule
\end{tabular}}%
\label{viqa} 
\end{table}%

\section{Limitations and Future Work}
Although our proposed DRI-IQA model can effectively enhance the performance on IQA tasks, some areas can be improved. For instance, in the restoration network, we only validate the effectiveness of NAFNet. In future work, we may conduct experiments on GAN-based restoration networks \cite{gan1, gan2} or diffusion-based networks, which have better visual effects. 

Additionally, we plan to conduct transferability experiments on multiple IQA models other than MANIQA. Moreover, the representation can be pre-trained on a larger scale high quality dataset to achieve higher metrics and better generalizability.

\section{Conclusion}
In this paper, we propose the Dual-Representation Interaction method for No-Reference Image Quality Assessment (DRI-IQA). Thanks to degradation representation and the assistance of the restoration network, DRI-IQA is appropriately applied to the IQA task for both restored and authentically distorted images. The DRI-IQA method utilizes only one encoder to generate quality representation and degradation representation simultaneously. In a dual-representation manner, the restoration assistance and Representation-based Semantic Loss (RS Loss) cooperatively increase the interaction between the two halves of the feature. Experiment results on four standard datasets demonstrate the outstanding performance and effectiveness of DRI-IQA. 

Moreover, the dual-representation interaction (DRI) method has great potential to be applied to other vision tasks and integrated with current SOTA methods to boost their performance and generalization ability.

{\flushleft \textbf{Acknowledgement}} This work was supported by the National Natural Science Foundation of China under Grant 62171304 and partly by the Natural Science Foundation of Sichuan Province under Grant 2024NSFSC1423, and Cooperation Science and Technology Project of Sichuan University and Dazhou City under Grant 2022CDDZ-09.
62171304.


{\small
\bibliographystyle{ieee_fullname}
\bibliography{egbib}
}

\end{document}